# Title:
# Impact of a split injection strategy on mixing, ignition and combustion behavior in Premixed Charge Compression Ignition combustion

## Authors:
**Ulrich Doll[1]\*, Christophe Barro[2,3], Michele Todino[4], Konstantinos Boulouchos[2]**

1. Paul Scherrer Institute, Experimental Thermal-Hydraulics Group, Forschungstrasse 111, CH-5232 Villigen PSI, Switzerland
2. ETH Zürich, Institute for Energy Technology, Laboratory for Aerothermochemistry and Combustion Systems, Sonneggstrasse 3, CH-8092 Zürich, Switzerland
3. Vir2sense GmbH, Sihlbruggstrasse 109, 6340 Baar, Switzerland
4. Istituto Motori, IM-CNR, Viale Marconi, Napoli, Italy

\*Corresponding author: ulrich.doll@psi.ch

## Abstract
Mixing, ignition and combustion behavior in a rapid compression and expansion machine operated under Premixed Charge Compression Ignition (PCCI) relevant conditions are investigated by combined passive optical and laser-optical high-speed diagnostics. The PCCI concept is realized using a split injection schedule consisting of a long base load injection and two closely separated short injections near top dead center. Previous studies of close-coupled double injections under constant ambient conditions showed an increased penetration rate of the subsequent fuel spray. However, the aerodynamic gain from the preceding injection is counteracted by the density rise during the compression stroke under transient engine conditions. The study confirms that the rate of mixing of the subsequent fuel spray is significantly increased. Regarding combustion behavior, the thermodynamic analysis exhibits contributions of low temperature oxidation reactions of more than 20 % to the total heat release, with a notable amount of unburnt fuel mass varying from 25 to 61 %. The analysis of the optical data reveals the multi-dimensional impact of changes in operating parameters on the local mixture field and ignition dynamics. The onset of low temperature reactivity of the first short injection is found to be dominated by the operating strategy, while the location is strongly related to the local mixing state. Low temperature ignition of the consecutive fuel spray is significantly promoted, when upstream low temperature reactivity of the preceding injection is sustained. Likewise, it is shown that high temperature ignition is accelerated by the entrainment of persistent upstream low temperature reactivity.

## Highlights:
- Characterization of mixing, ignition and combustion in an RCEM operated under PCCI conditions using a split injection strategy with high-speed optical and laser-optical diagnostics.
- The momentum gain from the preceding jet is counteracted by the density rise during the compression stroke, leading to shorter penetrations of the consecutive jet, however, with an increased mixing rate.
- Small changes in the operating strategy have significant influence on the local mixing state and reaction kinetics.
- The location of low temperature ignition is strongly related to the local mixing state. Upstream low temperature reactivity significantly accelerates the transition to high temperature ignition.

# 1 Introduction

The need both to reduce CO$_2$ emissions as well as to meet aggravated pollutant emission targets poses significant challenges to the mobility sector. In this context, low temperature combustion (LTC) concepts for direct-injection diesel engines are of great interest, since they have the potential to increase overall engine efficiency while at the same time reducing pollutant NO$_x$ and soot emissions [1]. In contrast to the mixing controlled combustion in conventional diesel engines (where the process limiting rate is dominated by mixing rather than reaction kinetics), promising LTC concepts such as Homogeneous Charge Compression Ignition (HCCI) employ extensive premixing to create a homogeneous lean fuel-air cylinder charge. This charge then simultaneously auto-ignites at multiple spots towards the end of the compression stroke, resulting in faster heat release and pressure rise at significantly lower peak temperatures [2]. Being beneficial in terms of emissions and efficiency, however, a potential drawback of the HCCI concept lies in the challenging control of the combustion process. Other than in fuel mixing controlled conventional diesel or spark-ignited gasoline combustion, the HCCI rate limiting process is reaction kinetics, which are influenced by a complex interplay of mixture preparation and temperature control of ignition timing through intake air heating, exhaust gas recirculation (EGR) or exhaust valve closing timing [1, 3-6]. In this context, hybrid combustion concepts such as Premixed Charge Compression Ignition (PCCI) offer a huge potential to combine high efficiency/low emission characteristics of HCCI with the superior operability of conventional diesel combustion [5]. In PCCI, the rate limiting processes from conventional diesel (mixing) and HCCI (reaction kinetics) are approaching similar levels. This can either be achieved by the choice of the fuel, which affects the reaction kinetics or by the choice of the injection strategy, which affects the mixing. The current work focuses on the injection strategy.

The PCCI combustion strategy involves fuel injection early in the compression stroke allowing for a certain degree of in-cylinder premixing in order to keep combustion temperatures low and, thereby, avoiding NO$_x$ and soot formation. In general, ignition control in PCCI is achieved through high EGR rates [7] or intake condition variations such as varying temperature, pressure or fuel reactivity by blending with low cetane fuels [5]. In addition to above control mechanisms influencing combustion chemistry and in contrast to the HCCI approach, PCCI implementations retain the coupling between start of injection and start of combustion [5, 8]. The latter is especially attractive since it offers the opportunity to use standard injection systems instead of specialized hardware. The current study aims at investigating the mixing and ignition fundamentals of PCCI combustion in applying multiple injections.

Fuel split by multiple injections is an established strategy to reduce pollutant emissions in diesel combustion while increasing fuel economy [9-12]. Additionally, multiple injections can be applied to tailor fuel-air mixing [13-15] and ignition timing [14, 15] as well as to reduce unburned hydrocarbons (UHC) and CO by preventing fuel wetting and flame quenching at the piston wall [14, 16, 17]. In [18] the reaction kinetics using various fuels under HCCI conditions were studied and modelled. The effect of mixing various injection durations was investigated in [19]. It was shown, that the role of low temperature combustion is increasingly important with shorter injection duration. In the PCCI concept, the role of detailed reaction kinetics separating low and high temperature combustion effects in combination with the mixing behavior of short injections, embedded in a split injection schedule are expected to be important. However, understanding of the complex interplay between the relevant operational parameters, fuel-air mixing and combustion chemistry affecting the PCCI combustion process is still scarce and optical diagnostics play a crucial role in providing detailed spatially and temporally resolved insights into the underlying mechanisms. Bruneaux and Maligne [20] carried out a detailed study of the mixing and combustion behavior of short double diesel injections in a constant volume cell (CVC). The authors observed significantly enhanced mixing at the spray tip of the second injection caused by the flow field of the preceding injection. In addition, it was found that low-temperature combustion intermediates of the first injection were entrained by the subsequent jet, leading to an accelerated high-temperature ignition process. In [21], the study of double injections of n-dodecane by Schlieren imaging and planar laser induced fluorescence (PLIF) of formaldehyde within a CVC confirmed faster penetration and mixing rates of the second jet by entering a kind of "slipstream", thus benefiting from the sustained momentum and turbulence of the previous injection. In support of the findings of [20, 22], the authors conclude that the ignition delay of the second jet is significantly reduced by the interaction with cold-flame reaction products of the first injection. At varying ambient temperatures from 900 K to 750 K, combustion efficiency was reduced from 95 % to 75 % and a persistent formaldehyde PLIF signal was associated with an incomplete high-temperature combustion process. By investigating the spray tip penetration of double injection schemes in an optical two-stroke engine using Schlieren diagnostics, Desantes et al. [23] confirmed the "slipstream" hypothesis and found a distinct relationship of faster penetration rates and shorter dwell times between injections as well as reduced ignition delays of the second compared to the preceding jet.

Even though above works provided valuable insights into how multiple injections influence mixing and combustion processes, discussed results were predominantly obtained at constant ambient densities and, therefore, lack the transient behavior of a full engine cycle. In addition, the studies focus on the interaction between the consecutive injections or mixing behavior of single injections rather than providing necessary operational characteristics for PCCI strategies such as mixture preparation. In this context, the current study aims at providing detailed insight into mixing and combustion behavior within an optically accessible rapid compression and expansion machine (RCEM) operated on a split injection PCCI strategy by means of passive optical and laser-optical high-speed diagnostics. In a first measurement campaign, ignition and combustion were characterized by performing combined $OH^*$ chemiluminescence, Schlieren and formaldehyde ($CH_2O$) planar laser induced fluorescence (PLIF) measurements. While from the first appearance of $OH^*$ kernels the onset of high-temperature (HT) combustion is determined, $CH_2O$ serves as an indicator for preceding low temperature (LT) fuel oxidation reactions. In a second experimental campaign, TMPD[1] tracer PLIF under inert conditions was used to study the effect of applying multiple injections on pre-ignition fuel-air-mixing. In order to identify a pertinent liquid core and the boundaries of the fuel spray, tracer PLIF was combined with Mie scattering and an additional Schlieren setup. The optical data is used to extract global jet metrics, to explore the influence of varying operational parameters on the in-plane mixing state and to discuss the resulting ignition and combustion dynamics. The analysis of optical data is complemented by a comparison of the jet metrics with literature correlations as well as a comprehensive thermodynamic analysis of global heat release rates and fuel consumption, finally linking the observed findings to engine operation and efficiency.

## 2 Experimental setup and evaluation methodology

### 2.1 Rapid compression and expansion machine

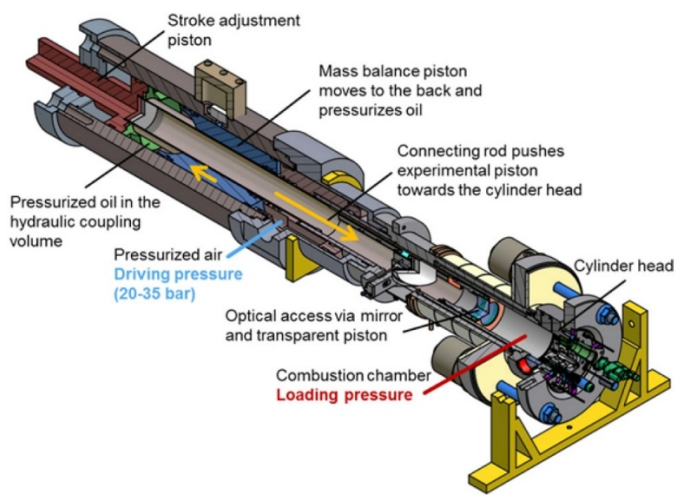

**Fig. 1: Schematic view of the RCEM: Pressurized air moves the mass balance piston towards the stroke adjustment piston, accelerating the working piston towards the cylinder head.**

The RCEM is a free-floating piston test rig (Fig. 1), originally designed by Testem and further modified in-house (i.e. equipped with a liner heating and modified piston and cylinder head(s)). The RCEM provides excellent optical access and very flexible operating parameters such as compression ratio, charge pressure and temperature. Its modular design allows the custom design of different cylinder heads, which can allows different accesses for diagnostic equipment, windows and injectors depending on the needs of the particular experimental campaign. The RCEM used herein is capable of performing a compression and expansion cycle similar to an engine. The working principle of the RCEM is illustrated in Fig. 1, left. Pressurized air moves the mass balance piston, which in turn pushes the working piston towards the cylinder head. The free-floating working piston is accelerated until it approaches top dead center (TDC) and, thereby, compresses the combustible fuel-air mixture. Expanding gases in the combustion chamber are then forcing the working piston back towards bottom dead center (BDC). Both, TDC and BDC are depending on the driving pressure, the charge pressure and the converted energy as well as the combustion strategy. The machine's main bore has a diameter of 84 mm, the maximum stroke can be varied and was set to 245 mm. A solenoid injector with a 100 µm single conical coaxial nozzle was mounted on the cylinder periphery (5 mm below the flat cylinder head, 42.5 mm from the cylinder axis and inclined 5° towards the piston) [24]. The cylinder pressure is obtained with a combination of an absolute pressure sensor in a switch adapter (recording the pressure until 5 bar) and a relative pressure sensor which is pegged during the beginning of the cycle. The cylinder pressure and the piston position are recorded at a rate of 100 kHz during operation. Optical accessibility to the combustion chamber is realized by means of a mirror within the slotted shaft of the working piston (a transparent flat piston bowl with 4 mm depth), a window at the cylinder head (diameter 52 mm) and another window with 40 mm in diameter at the side. A more detailed description of the machine and further operational characteristics can be found in [25-28].

---

[1] n,n,n',n'-tetramethyl-1,4-phenylenediamine

## 2.2 Operating strategy and heat release rate analysis

In order to achieve the desired conditions for PCCI operation with multiple injections, it was found that the most suitable strategy consisted of a comparatively long first injection to provide the necessary base mixture, followed by two short pulses closely spaced to TDC. The operating condition have been selected in order to study cases with different states of mixing at ignition and simultaneously respect the operating limits of the machine. The main goals were to avoid high temperature combustion before the last injection was partially premixed (to avoid mixing controlled combustion, i.e. conventional diesel combustion) as well as to remain within the visible domain, so that mixing and ignition could be observed by the optical diagnostics. N-heptane was selected as mono component surrogate diesel fuel, because of their similar cetane numbers. The fuel injection pressure was set to 600 bar for all operating points (OPs). In case of the non-reactive measurements, the liquid fuel was blended with the TMPD tracer substance at a concentration of 1.5 g/l.

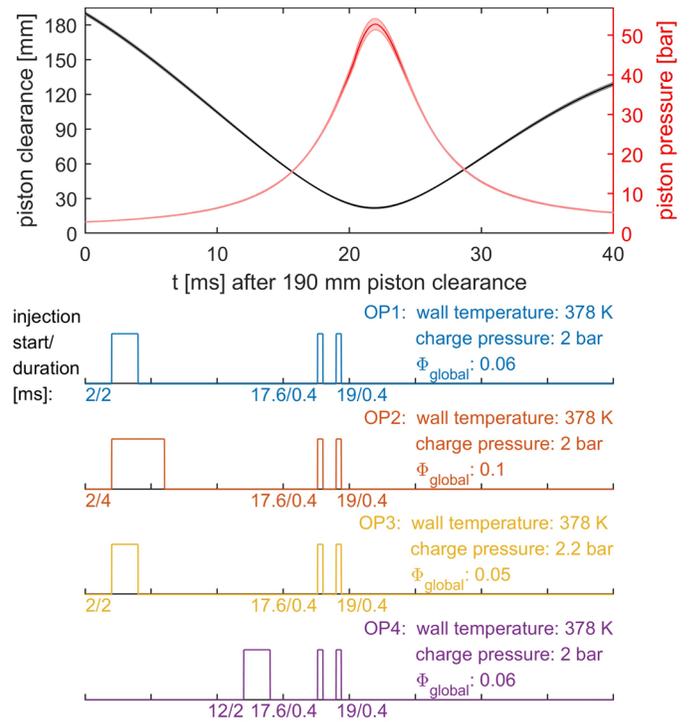

Fig. 2: (Top) Ensemble averaged time traces of piston clearance and pressure at reactive OP1. Shaded areas denote cycle-to-cycle variations of the RCEM. (Bottom) Injection schedules for investigated OPs: Trigger pulses indicate injector energizing times with respect to the start trigger event at 190 mm piston clearance. Respective timings (start of injection/duration) are stated below. Further characteristics for each OP are summarized on the right.

The resulting operating strategy is summarized in Fig. 2. All injection sequences as well as the optical data acquisition were started by a trigger event initiated during the compression stroke as soon as the accelerating working piston reached a distance of 190 mm with respect to the cylinder head. Unless stated otherwise, time axes within the manuscript are referring to that instance in time. Ten repetitions at each OP were carried out. The top panel of Fig. 2 shows representative time traces of ensemble averaged piston pressure and piston clearance at reactive OP1. Below, the injection schedules and further operational details of the four investigated OPs are summarized. Piston wall temperature and charge pressure were the same for OPs 1, 2 and 4, whereas OP3 featured a slightly elevated charge pressure resulting in a reduced global equivalence ratio. In case of OPs 1, 3 and 4 the same amount of fuel was injected, while for OP2 the energizing time of the first injection was doubled to increase the base load. For OPs 1, 2 and 3, the first base load injection occurred early in the compression stroke, allowing for a prolonged dwell aiming at enhanced premixing, while for OP4 the first injection was delayed by 10 ms. Timing and duration of second and third injections were selected to be the same for all OPs. It has to be noted that due to the free-floating piston design of the RCEM, above modifications to operational characteristics affected the overall compression ratio, which was 10.5 for OPs 1 and 4, 9.9 for OP2 and 8.7 for OP4, respectively.

The thermodynamic analysis has been performed with an in-house written software [29], which has already been used in previous works [27, 28, 30]. From the in-cylinder pressure signal and an estimation for heat transfer, the heat release rate has been obtained.

## 2.3 Optical diagnostics

### 2.3.1 Instrumentation

As indicated above, two successive test campaigns were carried out. In a reactive study, the effect of multiple injections on ignition and combustion behavior were investigated while in a non-reactive campaign, the impact of the split injection schedule on fuel-air mixing was explored. The first setup combined Schlieren, OH* chemiluminescence as well as $CH_2O$ PLIF imaging, the second Schlieren, Mie scattering as well as tracer (TMPD) PLIF measurements. All measurements were performed out at 10 kHz repetition rate.

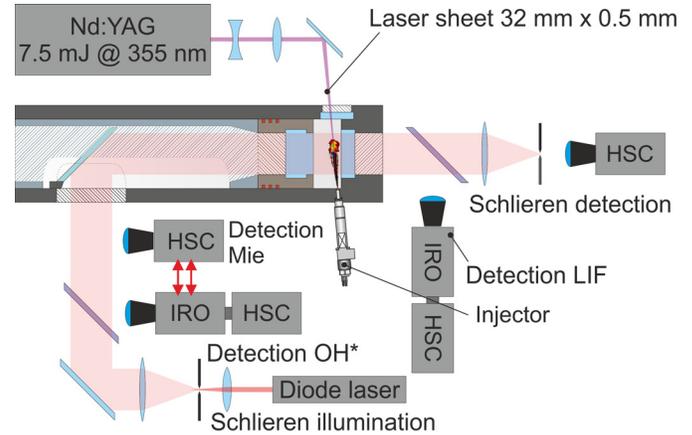

Fig. 3: Optical setup for simultaneous high speed CH2O-PLIF, Schlieren, and OH*-chemiluminescence imaging (reactive) and TMPD tracer PLIF, Schlieren, and Mie scattering imaging (non-reactive)

Fig. 3 shows a schematic drawing of the optical arrangement. The same Schlieren setup was applied for both reactive and non-reactive campaigns. Schlieren illumination was implemented based on a Cavitar CaviLux Smart diode laser (690 nm, pulsewidth 1 µs). The laser output was collimated by a combination of an aspherical (2", f=40 mm) and a spherical lens (4", f=400 mm). A 4" bending mirror was used to direct the collimated beam into the RCEM. Behind the machine, a second spherical lens (4", f=400 mm) was placed to focus the outgoing laser light onto a circular Schlieren cutoff aperture (diameter ~1 mm). Schlieren images were recorded using a LaVision HSSX camera equipped with a Nikkor VIS lens (f=105 mm, f/2.8).

Both formaldehyde as well as TMPD PLIF signals were excited using the third harmonic output of an Edgewave IS400 high-speed laser at 355 nm. Operated at 10 kHz repetition rate, the laser's output power was about 75 W with a pulse width of 6.5 ns. The laser beam was formed into a collimated sheet of 32 mm height using a combination of cylindrical lenses (f=-25 mm, f=250 mm) and a diaphragm. A 2" dichroic mirror was used to direct the light sheet into the RCEM through the lateral window. The beam waist was adjusted by a third cylindrical lens (f=1000 mm), resulting in a sheet width < 1 mm throughout the visible domain. As indicated in Fig. 3, the light sheet was slightly tilted to match the 5° injector inclination. Since the fuel spray is pushed towards the piston head during the compression stroke, Srna et al. [31] found that the effective symmetry axis of the spray cone has an angle of 3.5°, which was adopted for the arrangement of the light sheet in the present experiments. In both setups, LIF signals were directed towards a LaVision HSS6 high speed camera equipped with a high speed image intensifier (HS-IRO, gate width 150 ns, gain 62) using the same dichroic mirror (4", HR 45 300-450 nm). For $CH_2O$ LIF detection, a Nikkor VIS lens (f=105 mm, f/2.8) and a Hebo color glass filter (B05) were mounted, while in case of TMPD LIF a Sodern Cerco UV lens (f=100 mm, f/2.8) combined with a bandpass filter (Edmund optics, 400±25 nm) were used.

OH* chemiluminescence was recorded by means of a LaVision HSS6 high speed camera with high speed image intensifier HS-IRO (gate width 75 µs, gain 60), equipped with a Sodern Cerco UV-lens (f=100 mm, f/2.8) and a bk Interferenzoptik bandpass filter (313±7 nm). A dichroic beamsplitter (Eksma, 4", HR 45 305-320 nm) was used to deflect the back-scattered OH* signal towards the detection system while the Schlieren laser illumination passed. For the non-reactive tests, in order to detect Mie scattering from the liquid fuel spray excited by the Edgewave laser, the image intensifier was removed and the camera lens as well as the bandpass filter were replaced (lens: Coastal optics, f=105 mm, f/4.5; filter: bk Interferenzoptik, 355±10 nm). Additionally, a different dichroic beamsplitter (CVI, 4", HR 45 351 nm) was mounted to deflect the back-scattered Mie signal.

### 2.3.2 Data acquisition and evaluation of optical data

For both reactive and non-reactive campaigns, the start of image acquisition was externally (position) triggered when the working piston reached a distance of 190 mm to the piston head. A second position trigger was used to activate the image intensifier at 60 mm to prevent over-exposure of the intensifier caused by highly fluorescent liquid fuel components pertinent in the 1st injection. At an acquisition rate of 10 kHz, each acquisition channel acquired 400 images and 10 repetitions were carried out at each OP. Since the study focuses on pre-ignition mixing of 2nd and 3rd injection and the combustion behavior near TDC, image processing is limited to time steps 15 ms to 30 ms after the initial trigger event.

In a first evaluation step, all acquired images are mapped to a common grid defined by a transparent dot-pattern target using the LaVision Davis software's built in image dewarping and mapping routines. The origin refers to the injector nozzle tip, the resulting spatial resolution is 0.075 mm/pix. All images are multiplied with a circular mask to account for the viewing windows diameter of 52 mm.

OH* chemiluminescence measurements are used to asses the onset of auto-ignition as well as the duration of the HT combustion event. Images are background corrected and Gaussian smoothing is applied. Mie scattering data has been carefully examined and it has been found that the core of the liquid jet does not enter the visible range in any operating condition. Therefore, no Mie scattering results are presented. In the following, this section will focus on the more complex evaluation of Schlieren, $CH_2O$ PLIF and tracer PLIF optical data.

*Schlieren image processing*
In a first step, Schlieren images are flat-field corrected. The flat-field is calculated by averaging the first ten images of each Schlieren image series, whereby the Schlieren signal is still unaffected by the flow and therefore intensity variations exclusively result from the optical setup. In order to extract the contours of the fuel jet, images were further processed by applying an adaptive background subtraction method [32, 33]: The image acquired just before the fuel jet enters the visible area serves as a background image and is subtracted from the subsequent frame. The contrast of the jet area in the background corrected image is further enhanced by applying two-dimensional median filtering, threshold filtering and automated removal of remaining spurious objects, so that eventually the spray contours are obtained as the boundary of the fuel spray area. Jet penetration length and spray angle are then extracted from the spray contours. Finally, the original background image is modified by replacing all image regions outside the jet with corresponding areas from the current frame, while maintaining intensities within the jet boundaries from the previous image. The adapted background image then serves as new background for the next image.

*$CH_2O$ PLIF image processing*
$CH_2O$ PLIF images suffer from background structures related to spurious light from the Schlieren laser, which is partially reflected by the dichroic beamsplitter. Other background sources involve secondary reflections of the LIF signal and fluorescence exited at the injector head, both illuminating the moving working piston surface. The background is corrected by scaling and subtracting an average background image generated from multiple frames near the end of the acquisition cycle when the formaldehyde fluorescence has disappeared. A second circular mask of 36 mm in diameter is used to remove for remaining piston head artifacts. Finally, Gaussian smoothing and threshold filtering are applied to each image.

*Tracer PLIF image processing and transformation to equivalence ratio*
The evaluation of the tracer PLIF images is initiated by several image processing steps beginning with moderate Gaussian smoothing of the raw data images. Similar to $CH_2O$ PLIF images, the excited tracer fluorescence causes background artifacts by secondary fluorescence scattering off surfaces. These contributions vary in position and intensity from image to image as they primarily originate from fluorescence emitted by the moving fuel jet. In spatially averaging the current frame in radial direction, an intensity profile reflecting intensity variations along the spray axis is produced. This profile is then scaled and subtracted from each line of the current frame in a trial-and-error procedure, so that erroneous background intensities vanish. A second background artifact arises from a direct reflection of the tracer fluorescence emitted by the fuel jet off the working piston's glass surface. This mirrored portion of the jet is corrected by subtracting an axially shifted and scaled representation of the actual image.

The quantitative interpretation of the processed tracer PLIF images as equivalence ratio fields relies on the procedures outlined in [26, 34, 35]. The interpretation of measured tracer PLIF intensities as equivalence ratio $\Phi$ is governed by the following equation

$$\Phi(x,r,t) = \frac{I(x,r,t)}{I_0(t)\, I_{\text{ff}}(x,r)} \cdot \frac{AFR_{\text{stoich}}}{\phi(T)\, K\, \rho_{\text{air}}(p,T,t)}, \tag{1}$$

where $I$ denotes the background corrected PLIF intensity per spatial resolution element $(x, r)$ at time step $t$. Pulse-to-pulse variations of the laser energy are represented by $I_0$ and were not recorded during the measurements, since low fluctuation amplitudes below 3 % (rms) were expected based on previous experiments [26]. Due to some unknown reason, the pulse energy varied by 20 % (rms) in the current experiments, which has significant impact onto the measurement accuracy as will be shown below. To account for the efficiency of the acquisition arrangement as well as the spatial distribution of the incident laser light, a flat-field correction is performed: By flooding the combustion chamber with low concentration fuel/tracer vapor and in assuming a homogenous vapor distribution, remaining intensity variations within the field of view can be attributed to the efficiency of the detection devices and the laser beam profile, the latter being assumed temporally constant. A flat-field image $I_{\text{ff}}$ is then generated by averaging over an ensemble of 1000 individual images.

While the first expression on the right side of Eq. (1) introduces image operations that are specific to the present setup, the second term contains parameters referring to the conversion to equivalence ratio. The temperature dependent TMPD fluorescence quantum yield is denoted by $\phi$ and is obtained from an empirical model based on available literature data [33, 36-38]. The bulk piston temperature $T$ as well as the air density $\rho_{air}$ at each time step are calculated from measured piston pressures $p$ in assuming adiabatic compression/expansion. In assuming rotational symmetry of the fuel spray, the proportionality constant $K$ links the injected fuel mass to the volume integrated intensity and is used to transform the quantum yield corrected fluorescence intensity to local fuel density [34, 35]. The operation inherently assumes a low tracer concentration so that absorption of laser energy can be neglected and that the excitation energy of the probed transition stays below the saturation limit. The image at 18.9 ms after the start trigger is selected to perform the procedure, since at this time step the fuel spray of the 2$^{nd}$ injection is visible to its full extend within the camera's field of view. The constant is then used to normalize the respective image series. Finally, the expression is multiplied by the stoichiometric air-fuel-ratio $AFR_{stoich}$ for n-heptane-air mixtures.

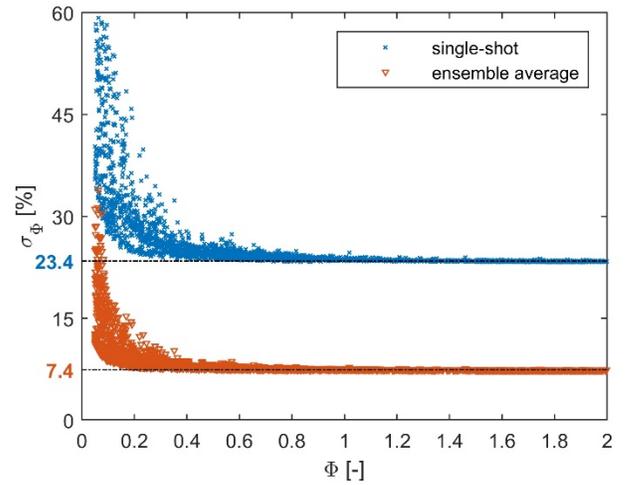

Fig. 4: Single-shot (blue, ✕) and ensemble averaged (red, ▽) precision of equivalence ratio.

Precision of the measured equivalence ratios is analyzed by applying the principle of propagation of uncertainty to Eq. (1). Relevant sources of uncertainty are camera noise, deviation of the fuel jet volume from rotational symmetry and the pulse-to-pulse variation of laser energy. Fig. 3 summarizes the relative precision analyzed for a representative field of instantaneous equivalence ratios, with each symbol representing a single pixel. For values of $\Phi < 0.3$, the precision is determined by camera noise and is, therefore, growing towards lower $\Phi$. At higher equivalence ratios an asymptotic decrease of the resulting precision can be observed, which reaches a minimum value of 23.4 % and is mainly dominated by the 20 % pulse-to-pulse variation of the laser energy. Ensemble averaging over ten repetitions increases the precision by a factor of ~3, converging towards 7.4 % for $\Phi > 0.3$.

## 3 Results and discussion

### 3.1 Jet metrics

The temporal evolution of the jet penetration length and the spray angle of the 2$^{nd}$ and the 3$^{rd}$ injection extracted from the Schlieren contours of the non-reactive measurement campaign are summarized in Fig. 5 (left). The injections are of short duration, so that both are completed before the respective fuel spray enters the visible area after ~0.2 ms. The penetration length develops almost identical for all OPs and for both injections. Differences between the curves are well within the experimental spread over the ten repetitions marked by the gray shaded area. The same holds true for the spray angles associated with the 2$^{nd}$ and the 3$^{rd}$ injection, which approach an average value between 21° and 22° when the full width of the spray jet becomes visible within the field of view after ~0.7 ms. From these results it is apparent that the parameter changes according to the operating strategy outlined above have no significant impact on the global jet metrics. In contrast to previously discussed results obtained from optical investigations of multiple injections in CVCs , no faster penetration of the subsequent (3$^{rd}$) injection is observed in our data. On the contrary, an opposing effect is evident in jet tip penetration differences depicted in Fig. 5 (top, right), leading to decreased penetration rates of the 3$^{rd}$ injection before 1 ms after el. SOI for all OPs. An explanation lies in the transient nature of RCEM ambient conditions compared to CVC experiments: Since the chamber pressure rises during the compression stroke, the momentum gain caused by the "slipstream" effect is counteracted by the associated density increase and, therefore, results in an overall reduced penetration length instead. At later stages, differences in jet penetration for OPs 2 and 4 stabilize around 1 mm, while in case of OPs 1 and 3 the 3$^{rd}$ injection gains on the preceding jet so that it eventually reaches similar penetration distances. This behavior can be explained by the deceleration of the working piston when approaching TDC, which leads to a significant decrease in density growth rate and, thereby, to a decreasing resistance against the expanding 3$^{rd}$ jet in relation to the expected momentum gain from the 2$^{nd}$ injection. While the penetration length exhibits a pronounced behavior in connection with the engine cycle, no conclusive trends can be deduced from the spray angle data.

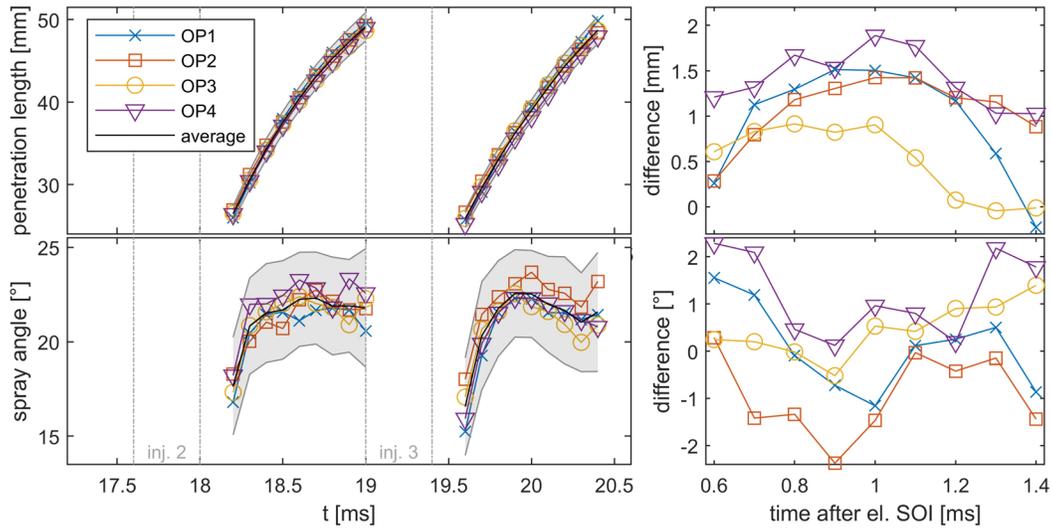

**Fig. 5:** (Left) Ensemble averaged jet penetration length (top) and spray angle (bottom). Black solid curves represent the respective average over all OPs, shaded areas indicates the average spread over the ten repetitions. Dashed-dotted vertical lines mark the energizing time of 2$^{nd}$ and 3$^{rd}$ injection. (Right) absolute difference of ensemble averaged jet penetration length (top) and spray angle (bottom) of 2$^{nd}$ minus 3$^{rd}$ injection.

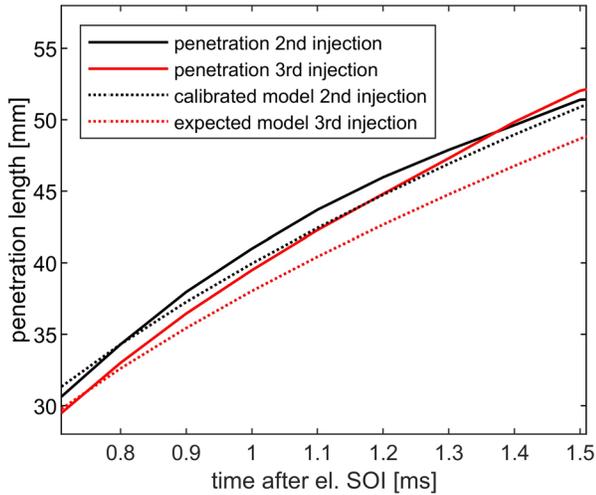

**Fig. 6:** Ensemble averaged jet penetration length for 2$^{nd}$ and 3$^{rd}$ injection compared with calibrated 2$^{nd}$ and expected 3$^{rd}$ injection derived from Hiroyasu's model for OP1.

Fig. 6 shows in solid the measured penetration length of the spray of OP1 (black 2$^{nd}$ injection, red 3$^{rd}$ injection, similar for the other OPs). The black dashed line indicates the penetration length from a spray model governed by the following equation

$$S = a_0 \sqrt{\sqrt{\frac{P_f - P_a}{\rho_a}} \cdot d_0 \cdot t}, \qquad (2)$$

where $S$ is the spray penetration length, $a_0$ is a parameter to be calibrated and the term under the double square root is the similar to the spray models from [39-41]. $P_f$ and $P_a$ are the fuel respectively the charge pressure, $\rho_a$ is the charge density, $d_0$ is the nozzle reference diameter and $t$ the time. The parameter $a_0$ has been calibrated to match the experimental data from the 2$^{nd}$ injection. The red dashed line uses the same model with the same parameter $a_0$, but with the charge pressure and density evolution from the 3$^{rd}$ injection. The graph shows, that the estimated penetration length for the 3$^{rd}$ injection (red, dashed) differs significantly from the experiment (red, solid). This behavior has already been reported in a constant volume chamber [21], where the penetration of the second of two consecutive sprays penetrates faster due to the discussed aerodynamic effects. However, it is important to note, that in an engine environment, even though the consecutive spray "benefits" from the preceding jet, the penetration does not necessarily need to be faster, since the consecutive spray is injected into a higher pressure and density environment as long as the injection timing is during the compression stroke.

## 3.2 Mixing dynamics

### 3.2.1 Local mixing phenomena

The section targets at revealing the complex local mixing dynamics between the 2$^{nd}$ and the 3$^{rd}$ injection and to establish a relation to the matrix of operating points. Owing to the limited precision of the single-shot equivalence ratio maps, the following analysis will rely on ensemble averaged results, denoted by $\langle \Phi \rangle$. Ensemble averaged equivalence ratio maps of 2$^{nd}$ and 3$^{rd}$ injections for OP1 are compared in Fig. 7 (left). A threshold of $\langle \Phi \rangle = 0.5$ is used to remove the base load contributions in to enhance the contrast of the jet boundaries.

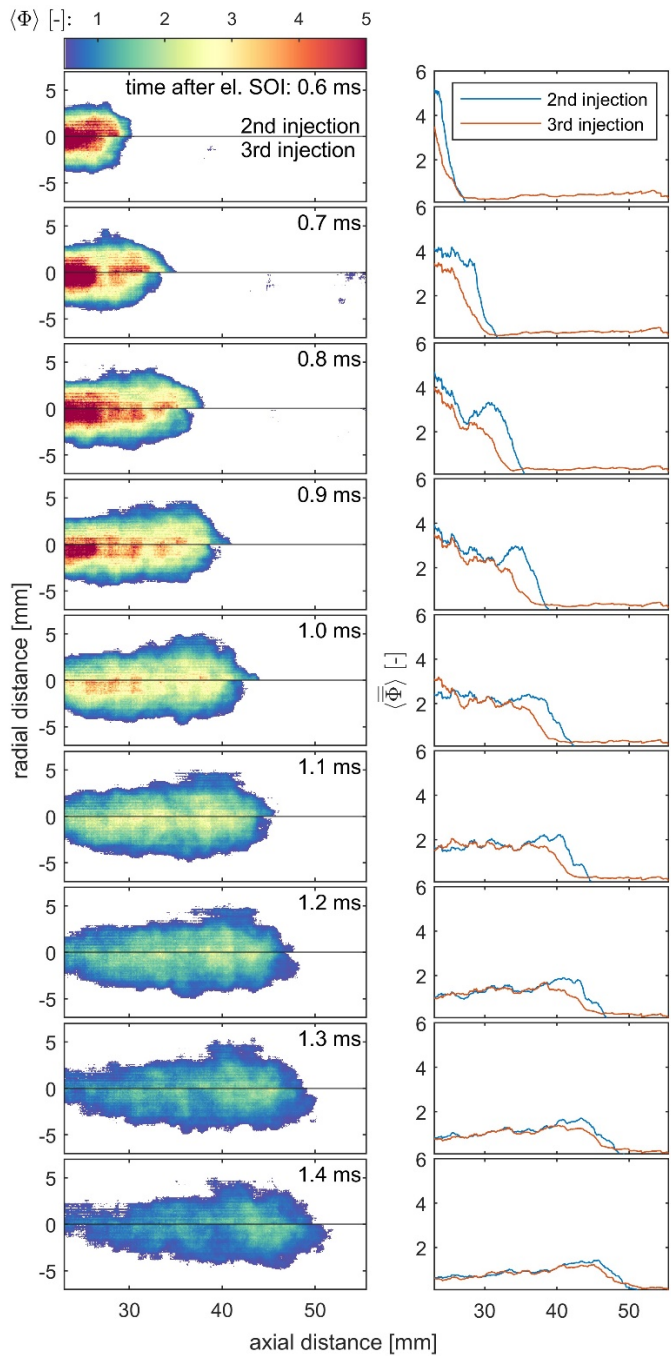

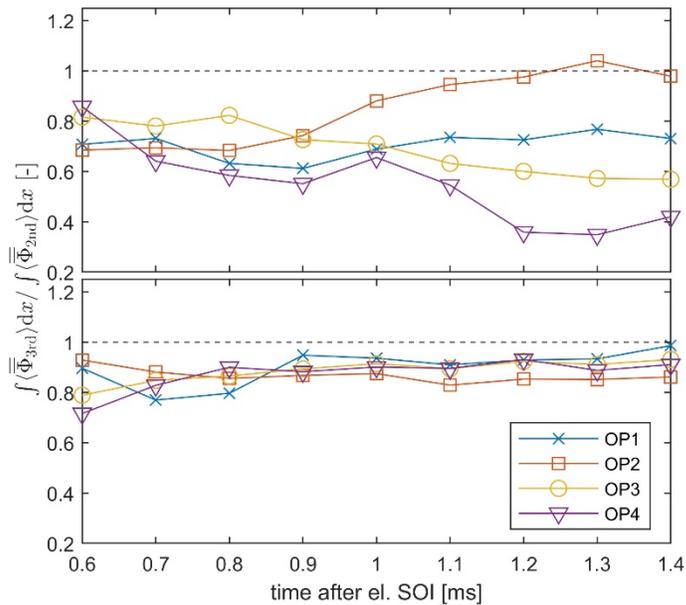

Fig. 8: Ratio of integrated ensemble and cross-sectional average equivalence ratio between 2nd and 3rd injection (top) at the spray head ($x_{tip} - 5$ mm) and (bottom) for the remaining visible spray body.

Fig. 7: (Left) Comparison of ensemble averaged equivalence ratio $\langle \Phi \rangle$ maps of 2nd (upper half) and 3rd (lower half) for several time steps after el. SOI. (Right) Corresponding ensemble and cross-sectional average equivalence ratio $\langle \overline{\overline{\Phi}} \rangle$ profiles.

The mixing fields confirm the faster penetration of the 2nd jet compared to the 3rd injection. Corresponding cross-sectional average equivalence ratio profiles in Fig. 7 (right) develop a steep gradient at the spray head of injection 2, which is related to the formation of a stagnation plane and is typical for pulsed jets emerging into a quiescent medium [21, 42]. In contrast, corresponding profiles of the 3rd injection exhibit a significantly smoother decay of the equivalence ratio at the spray tip, indicating enhanced mixing at the head of the post-injection due to the turbulence induced by the preceding jet [13, 20, 21].

The enhanced spray tip mixing of the 3rd fuel jet is further quantified by integrating the ensemble and cross-sectional averaged equivalence ratio of 2nd and 3rd injection at the spray head ($x_{tip} - 5$ mm) and by calculating the ratio between these quantities. Values below unity indicate a faster spray head mixing during the 3rd injection and vice versa. The resulting curves for all investigated operating conditions are summarized in Fig. 8 (top). Until 0.9 ms, ratios are within a range of 0.6 to 0.8, indicating a stable improved mixing rate at the spray head of the 3rd fuel jet for all OPs. While OP1 and OP3 ratios stay within that range for all time steps, OP2 values gradually grow until reaching unity beyond 1.3 ms, indicating similar spray head mixing rates of the 2nd and the 3rd injection. The OP4 curve instead begins to fall steadily after 1 ms and finally approaches a minimum value of about 0.4. This suggests a continuous increase of the spray head mixing rate of the 3rd injection.

As opposed to the markedly different spray head structure of the 2nd and the 3rd injection, the upstream distributions of the ensemble averaged equivalence ratio maps as shown in Fig. 7 (left) for OP1 are of similar shape and magnitude. The corresponding cross-sectional average equivalence ratio profiles confirm this observation by presenting comparable slopes and absolute values in the upstream regions. A comparison of the ratio of integrated cross-sectional average equivalence ratios of the remaining visible spray body in Fig. 8 (bottom) reveals an almost constant offset between 2nd and 3rd injection over all time steps. This leads to the conclusion, that the flow field induced by the 2nd injection results in a stable increase in the upstream mixing rate of the 3rd jet, independent of global parameter changes associated with the operating matrix.

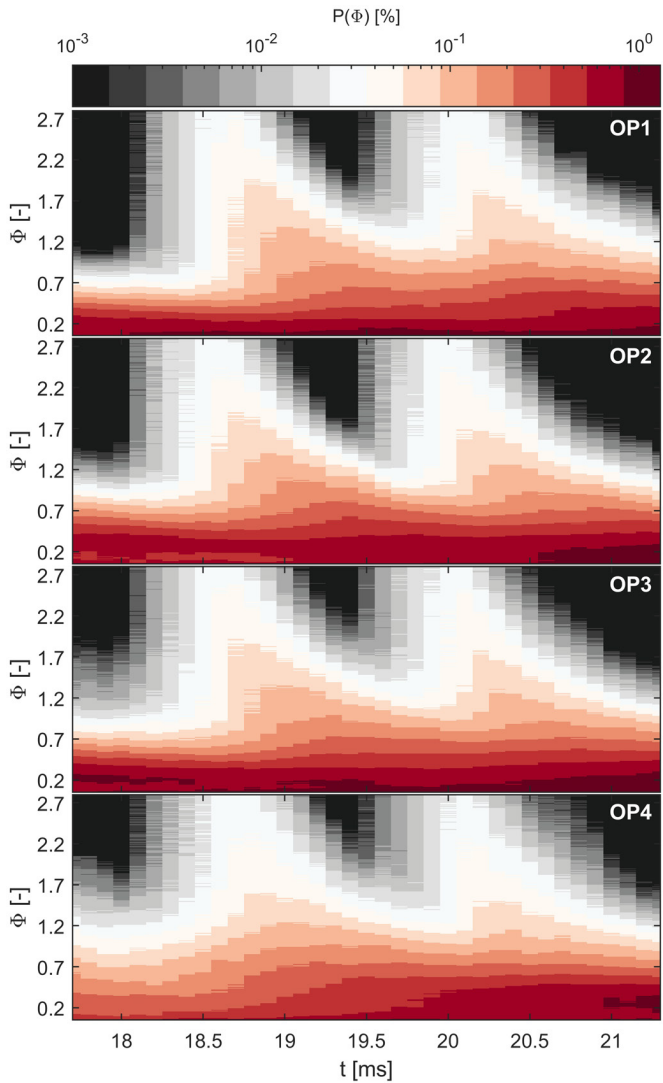

Fig. 9: PDF maps of the light sheet area: Each column represents an average PDF obtained by ensemble averaging the PDFs derived from single-shot equivalence ratio distributions at the corresponding time step. The probability of a certain equivalence ratio (y-axis) to occur at a specific time step (x-axis) is indicated by the color map in logarithmic scale.

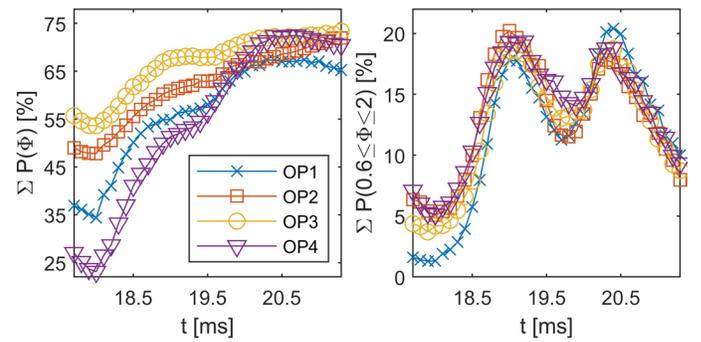

Fig. 10: Accumulated probabilities (excluding zero) $\sum P(\Phi)$ within the light sheet area over all $\Phi$ (left) and within the range $0.6 \leq \Phi \leq 2$ (right)

### 3.2.2 Global mixing state

To analyze the global mixing state, probability density functions (PDFs) from each single-shot equivalence ratio map are produced and ensemble averaged over the corresponding ten repetitions. The results of the analysis are summarized in Fig. 9. Each column of the displayed PDF maps represents the distribution of equivalence ratio within the area illuminated by the laser light sheet at a certain time step. The likelihood of a particular equivalence ratio to occur is indicated by the coloring.

At times before 18.1 ms, the distributions show the mixing state resulting from the base load (1st) injection, with all of them displaying a varying degree of stratification. The mixing at OP1 is the most advanced with a maximum probability around $\Phi = 0.2$ and no discernable contributions beyond $\Phi = 1.2$. OP2 exhibits a similar distribution, but slightly shifted to higher equivalence ratios due to the doubled duration of the base load injection. Even though being globally leaner, the mixing state prior to the 2nd injection of OP3 appears even more stratified with equivalence ratios exceeding $\Phi = 1.7$. The increase in stratification might be related to the higher charge pressure and the thereby reduced penetration depth, which changes the mixing dynamics of the base load injection. Finally, the base load stratification is most pronounced in the case of OP4, which is strongly related to the delayed 1st injection and the, therefore, considerably shortened time for fuel-air premixing. Starting with portions of the 2nd jet coming into view at 18.1 ms (0.5 ms after el. SOI), all maps are governed by a peak structure related to the 2nd injection. The steep rise at the beginning points to the wide range of equivalence ratio magnitudes contained in the initial jet. The mixing with the charge, driven by the momentum from the fuel injection leads to rapid decrease in equivalence ratio, resulting in a moderately stratified fuel-air mixture with absolute values below $\Phi = 2.2$. For time steps after 19.5 ms (0.5 ms after el. SOI), the 3rd injection starts to contribute accordingly. Finally, the in-plane mixing state approaches a similar degree of stratification for all OPs.

To further investigate characteristics of the global mixing state within the laser sheet area, the accumulated probabilities over all equivalence ratios are plotted against time and compared for all OPs in Fig. 10 (left). The resulting curves reveal, that only a fraction of the probed area is actually filled with fuel-air mixture and that there are pronounced differences between the OPs: While OPs 2 and 3 reach an initial filling state around 50 % and 55 % at 18 ms, respectively, OP1 reduces to 35 % of fuel-air mixture, whereas OP4 drops below 25 %. However, as time progresses, the double injection causes the light sheet plane to gradually fill up until all OPs reach a similar level around 65-70 % of the investigated area.

To give a first implication from the mixing state towards the onset of LT oxidation reactions, Fig. 10 (right) summarizes the accumulated probabilities in the range between $0.6 \leq \Phi \leq 2$ that are prone to initiate a LT combustion event [43]. In contrast to accumulating over all probabilities, confining the analysis to the specified equivalence ratio range results in similar distributions for all OPs. All curves start at a level around 5 % or below, which is followed by a steep growth caused by the contributions of the 2$^{nd}$ injection. After reaching a first maximum around 18.7 ms, leaning of the fuel-rich portions causes the accumulated probability to decrease again. The pattern repeats as soon as the 3$^{rd}$ injection enters the field of view, until the progression of mixing leaves 10-15 % of the global fuel-air mixture in a state that favors the onset of LT oxidation reactions.

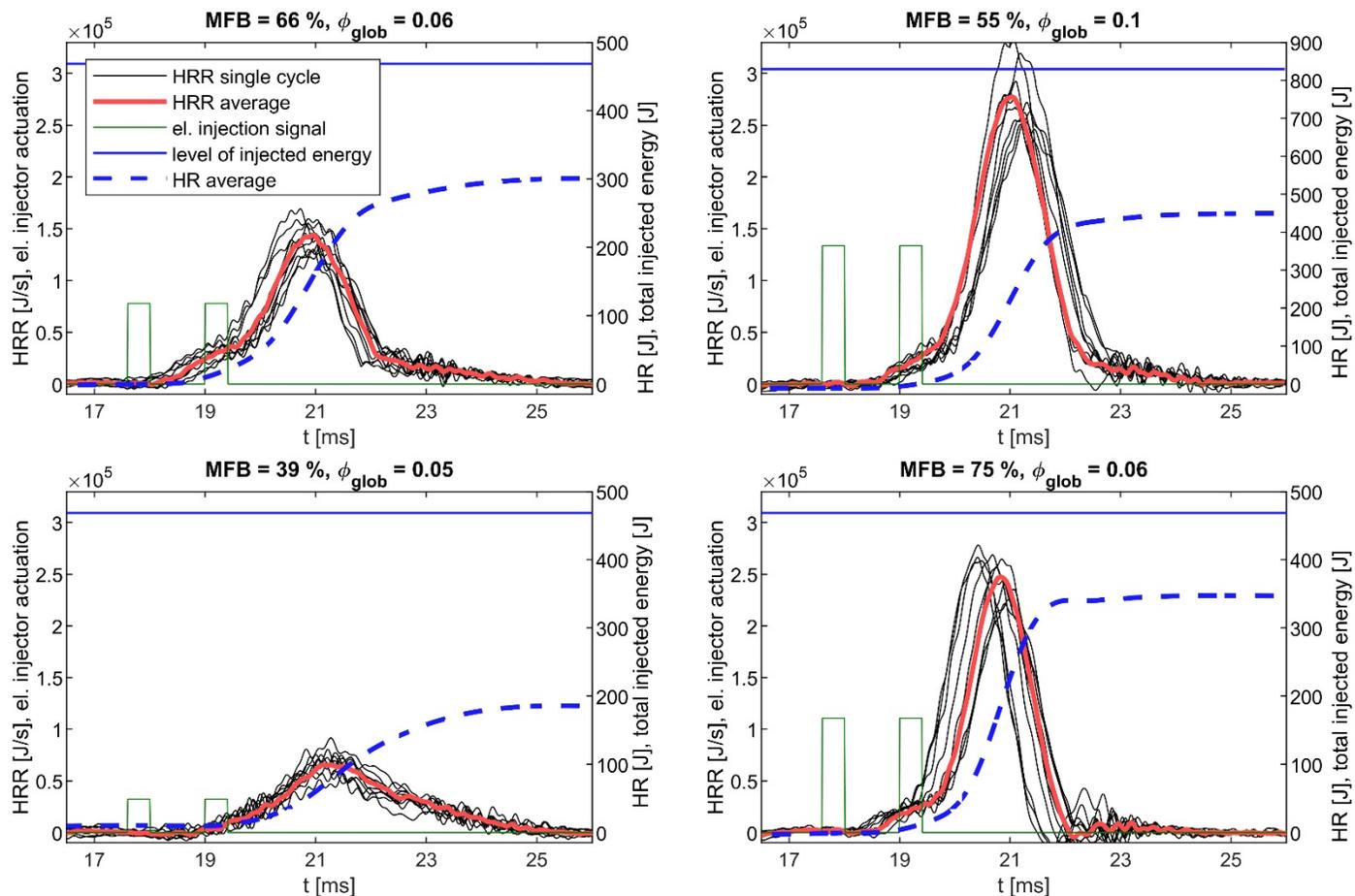

**Fig. 11:** Thermodynamic analysis: Heat release rate (black (single shots) and red (average)), electronic injection actuation signal (green, left vertical axis), cumulative heat released (blue dashed) and level of total injected energy (blue solid, right vertical axis). Top left: OP 1, top right OP 2, bottom left, OP3 and bottom right OP4.

Above findings reveal that, even though there are marked differences in the degree of premixing as well as fuel deposition after the 1$^{st}$ injection, all OPs approach a similar state of mixing after the 3$^{rd}$ injection. This leads to the conclusion, that the in-plane mixing state is primarily determined by the two post-injections, which are identical for all OPs, and not by the applied changes to the operating strategy. Since mixture preparation is essentially the same for all OPs, it is expected that differences in ignition and combustion behavior will be determined by local effects rather than by the global mixing state.

### 3.3 Combustion analysis

#### 3.3.1 Thermodynamic analysis and global combustion metrics

Fig. 11 displays the heat release rate (HRR), the total injected energy level and the converted energy for all four OPs. The graphs show trigger to the injector in green, the ten repetitions (black) and the average (red) of the heat release rate on the left vertical axes. In addition, the graphs show the injected energy level (solid) of all three injections and the evolution of heat released in dashed blue with the right hand vertical axes. The percentage of converted to injected energy as well as the global equivalence ratio is displayed in the according subfigure title. The visible 2$^{nd}$ and 3$^{rd}$ injection are always at the same time, with the same fuel pressure and the same duration. The operation conditions only vary in mass and timing of the 1$^{st}$ injection or charge air pressure (according to Fig. 2). Regarding OP1 as reference case, OP2 has a larger amount of fuel injected in the 1$^{st}$ injection, OP3 has a higher charge pressure and OP4 a later timing of the 1$^{st}$ injection. OP2, in comparison to OP1 shows a higher absolute amount of fuel burnt and a faster heat release rate. This is due to the richer base load after the 1$^{st}$ injection. The fuel mass from the 1$^{st}$ injection, which is entrained into the 2$^{nd}$ and 3$^{rd}$ spray and thus

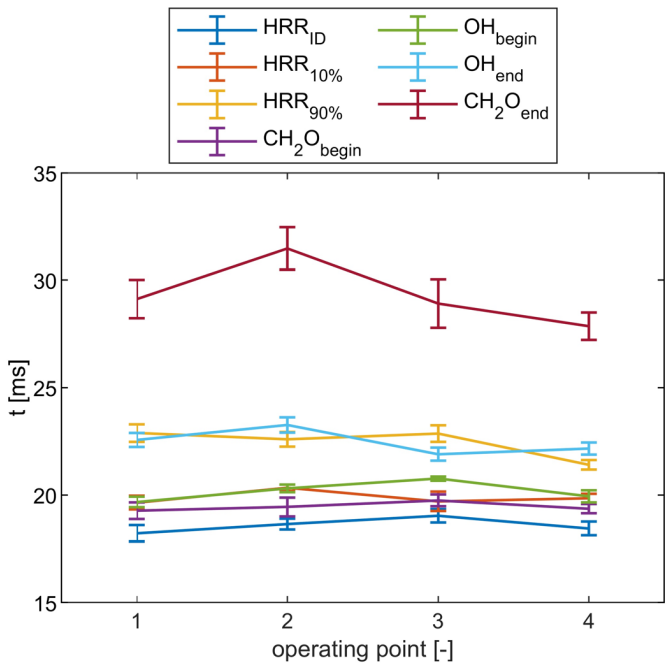

**Fig. 12:** Global combustion metrics: Ignition, 10% and 90% cumulative heat released (relative to the converted and not to the injected fuel, dark blue, orange and yellow), start and end of formaldehyde appearance (purple and red), start and end of OH* appearance (green, light blue)

forming a flammable mixture is larger. However, also the amount of fuel below the flammability limit is higher. Therefore, the relative amount of combusted fuel is reduced. OP3 in comparison to OP1 has a higher charge pressure and, therefore, a reduced global equivalence ratio. The further reduction of the equivalence ratio also reduces the regions with flammable mixtures and, therefore, the mass fraction burnt. In addition, this operating condition does not show a step in HRR as the other OPs. OP4 in comparison to OP1 has a later timing of the 1st injection, which means that the mixture at the start of the second injection is less homogeneous. This has the consequence that less fuel is entrained into the trailing edge of the second injection (Fig. 10, left). However, a mixture of richer charge is entrained at the tip of the second and third spray (Fig. 10, right), which eventually leads to a larger mass fraction burnt and a steeper HRR.

Fig. 12 compares the relevant timings from the thermodynamic analysis and the global optical metrics (i.e. start and end timing of the OH* signal the $CH_2O$ PLIF signal, representing formaldehyde). The figure shows in blue, orange and yellow the start, 10% and 90% point of the total heat released, in green and light blue the start and end of the OH* signal and in purple and red the start and end of the formaldehyde signal. The vertical axis is the time and the horizontal axis represents the four OPs. The thermodynamic begin of the HRR is the earliest detected value in all cases. The start of formaldehyde lies between start and 10% of the total heat released. The reason to see the two signals not on top of each other is that the formaldehyde signal is only visible inside the light sheet plane. The reactions outside this plane are not detected. In addition, the $CH_2O$ PLIF signal was contaminated with background artifacts, which, although corrected, decrease the signal-to-noise ratio and thereby raise the lower detection limit. On the other hand, the start of the OH* signal matches the 10% heat release for three of the four OPs. This means, that 10% of the HRR are already converted in LT combustion before the onset of HT combustion. For OPs 1, 2 and 4, the HRR in Fig. 11 also shows a significant change in rate at roughly 20 ms. The 90% point of the heat released matches quite well with the end of the OH* signal. However, the temperatures are still high and the available fuel under very lean conditions is still being converted in the LTC regime. The end of the formaldehyde signal is significantly later. For OPs 1, 2 and 3, the HRR in Fig. 11 is still visible after roughly 22 ms. For OPs 1 and 2, a significant change of rate is visible, whereas in OP3, the difference between LT and HT combustion is too weak. OP4 shows no significant HRR after 22 ms, and the formaldehyde signal ends first of all OPs. Additionally, OP4 is the operating condition with the highest fuel mass fraction burnt.

### 3.3.2 Ignition characteristics

Auto-ignition of high-pressure fuel sprays is a complex process that depends on the local history of the fuel-air mixture, combustion chemistry and the interaction between flow and reaction kinetics [43, 44]. In this context, global changes of operating parameters such as injection timing or charge pressure have a multi-dimensional effect on these local conditions, so that even small variations of the operating strategy can have a significant impact on the local ignition behavior. Additional operative degrees of freedom are introduced by the RCEM design used herein, where, due to the free-floating piston, the long 1st injection of OP2 as well as the increased charge pressure of OP3 impact the compression ratio and thus alter the temperature rise rates and peak temperatures during the compression stroke. Both the RCEM specific engine cycle as well as the local impact of global parameter changes find their expression within the global combustion metrics in Fig. 12. The following analysis aims at disentangling some of these effects.

The characterization of the ignition behavior in relation to the operating strategy relies on a combined analysis of high-speed Schlieren, $CH_2O$ PLIF and OH* chemiluminescence measurements. At that point it must be stressed that the following analysis is naturally restricted to the optically accessible part of the RCEM combustion chamber, which does not exclude HT or, in particular, LT oxidation reactions to occur outside the visible domain. Schlieren and OH* chemiluminescence images contain line-of-sight integrated information, whereas $CH_2O$ PLIF is localized to the laser light sheet plane, which, furthermore, does not exclude LT reactions to take place outside this plane but inside the visible area.

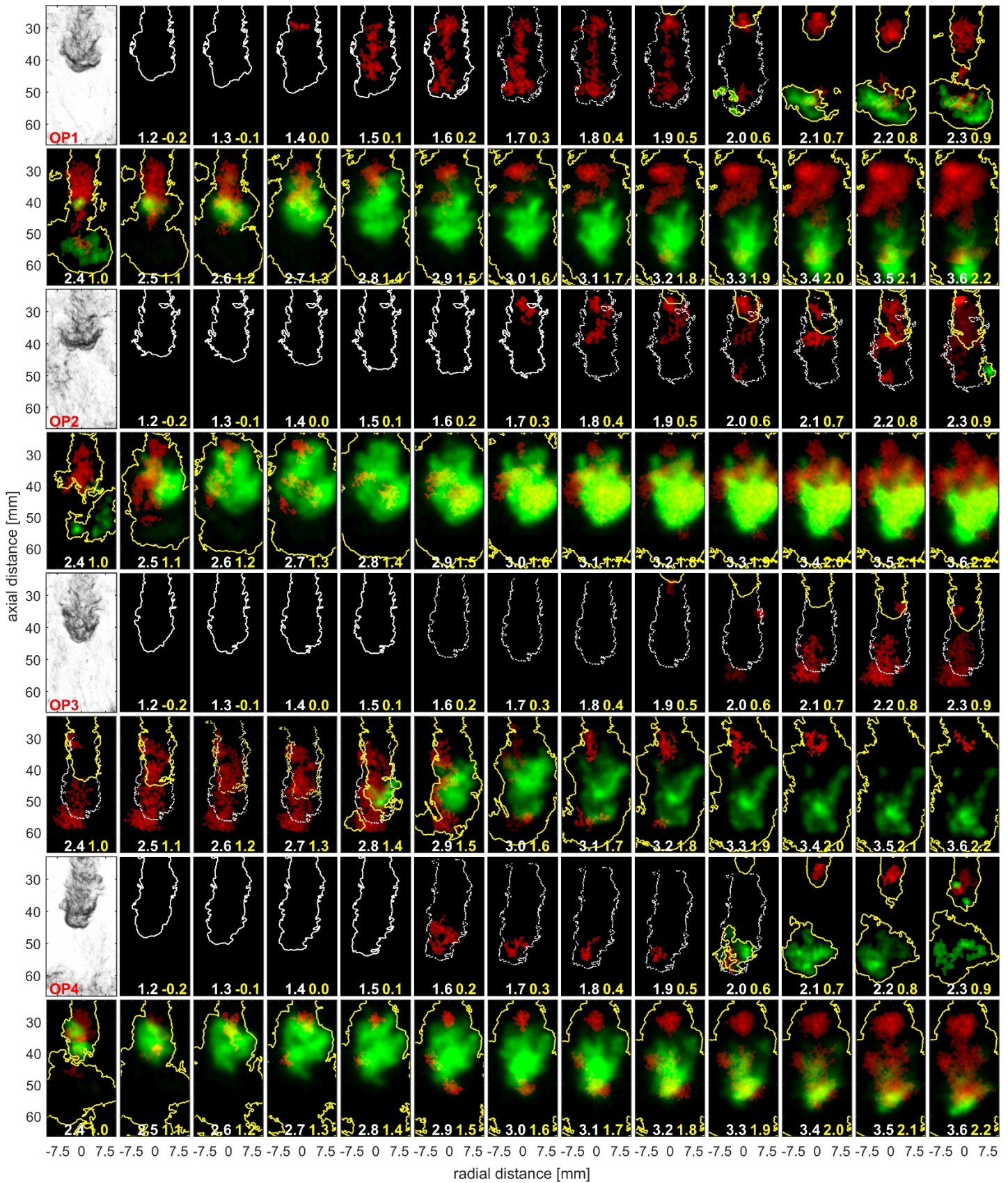

Fig. 13: Time-resolved ignition dynamics: Each OP occupies two rows and is introduced by a Schlieren image. Subsequent images contain the simultaneously acquired in-plane $CH_2O$ PLIF signal (red) and the line-of-sight integrated $OH^*$ chemiluminescence (green). Superimposed are the jet contours of 2nd (solid white) and 3rd injections (solid yellow) extracted from the simultaneously acquired Schlieren image data. Jet contours become dotted when tracking in the Schlieren data is no longer feasible. The last detectable contour is kept and translated along the jet axis by an amount related to the average jet tip velocity extracted from the non-reactive Schlieren data. Numbers at lower right denote the time after el. $SOI_2$ (white) or $SOI_3$ (yellow) in ms.

Page 13

Representative sequences of the ignition behavior for a single measurement run out of the ten repetitions for each OP are summarized in Fig. 13. Regarding OP1, a first spot of formaldehyde within the $2^{nd}$ fuel spray is formed 1.4 ms after el. $SOI_2$ about 30 mm downstream from the nozzle tip. At that time step, the mixing state appears to be ideal to promote LT ignition since the local equivalence ratio, due to fast mixing at the trailing edge of the spray results in higher temperatures then in richer conditions towards the spray tip [19] (Fig. 7, right). In addition, the global structure of the mixing field (Fig. 10, right) exhibits large portions of equivalence ratio ($\Phi < 2$) prone to undergo LT oxidation. The first LT ignition spot then grows rapidly in the downstream direction to finally cover the entire span of the $2^{nd}$ fuel jet at 1.8 ms. When the head of the $3^{rd}$ fuel spray comes into view at 1.9 ms after el. $SOI_2$, the LT reactivity related to the $2^{nd}$ injection already appears to decline, until at 2 ms only small portions at the spray head remain. At the same time, however, first HT ignition kernels appear at a similar axial location between 50 and 60 mm and merge into a single reaction zone, which then expands over several time steps, is transported downstream and disappears again 2.5 ms after el. $SOI_2$. In contrast to the $2^{nd}$ injection, the $3^{rd}$ fuel spray exhibits a distinctly different LT ignition behavior, as portions at the spray head immediately undergo LT oxidation reactions 0.6 ms after el. $SOI_3$. The shortened LT ignition delay between $2^{nd}$ and $3^{rd}$ injection agrees well with earlier observations and is related to the entrainment of hot LT reaction products from the preceding by the subsequent jet [20, 21]. In addition to this established explanation, the enhanced mixing at the head of the $3^{rd}$ spray due to the slipstream effect leads to leaner mixtures in comparison to the rich stagnation plane structure of the $2^{nd}$ jet (compare Fig. 7, right at 0.9 ms), which are more susceptible to undergo LT oxidation reactions. Formaldehyde fills the cross-section of the $3^{rd}$ jet, until at 1 ms after el. $SOI_3$, a HT ignition kernel related to the arrival of the $3^{rd}$ injection develops 40 mm downstream from the injector nozzle. In the following time steps, the HT ignition kernel expands in all directions and partially consumes the formaldehyde. The HT reaction zone reaches its maximum extend at 1.4 ms after el. $SOI_3$ and is then transported in downstream direction, whereby an intense formaldehyde signal develops in its wake. A similar observation of post-combustion formaldehyde was also made in [21] and is related to an incomplete consumption of the available fuel during the HT reaction. The higher intensity of the post-combustion $CH_2O$ PLIF signal, however, does not necessarily indicate a higher LT reactivity and should be interpreted with caution. Even though high pressure and temperature after HT combustion promote quenching of the formaldehyde fluorescence [45], this is counterbalanced by the absence of oxygen, which is found to have one of the highest quenching rates for formaldehyde [46]. In this context, since ambient conditions after HT combustion appear overall favorable to sustain LT reactivity, the remaining fuel partially undergoes LT oxidation reactions, which, however, do not proportionally contribute to the combustion efficiency (compare Fig. 11).

LT reactivity of OP2 starts about 0.3 ms after OP1, however, at a similar axial location within the boundaries of the $2^{nd}$ fuel spray. One time step later at 1.8 ms after el. $SOI_2$, the initial spot has grown in axial direction until it covers about half of the jet's cross-sectional area. In contrast to OP1, no LT reactivity occurs in the jet's downstream half until 2 ms after el. $SOI_2$, where a first formaldehyde spot emerges near the spray head. At 2.3 ms after el. $SOI_2$, LT oxidation reactions finally occupy a substantial part of the downstream half of the $2^{nd}$ fuel spray. The explanation both to the delayed onset of LT oxidation reactions as well as the postponed LT reactivity in the jet's lower half in comparison to OP1 is twofold: First, the lower compression ratio results in a reduced rise rate of the ambient temperature, so that it takes longer to trigger LT reactivity. Second, the higher base load leads to richer mixtures, such that in the lower half of the $2^{nd}$ spray the equivalence ratio is still too high (and temperature is too low) to undergo LT oxidation. In accordance with the time delay of 0.3 ms to the beginning of LT reactivity and at a similar axial location compared to OP1, a first HT ignition kernel is formed 2.3 ms after el. $SOI_2$ near the tip of the $2^{nd}$ fuel spray of OP2, spreads slightly and loses intensity rapidly in the following time steps. Other than OP1, the LT reactivity of the $2^{nd}$ fuel spray is still on-going when the $3^{rd}$ jet enters the visible domain 0.5 ms after el. $SOI_3$ and seems to be partially quenched at 0.7 ms due to colder portions of the $3^{rd}$ spray mixing with the LT combustion products of the $2^{nd}$ injection. The $3^{rd}$ fuel spray further expands and quickly fills with formaldehyde, until it reaches the outline of the HT reaction zone of the $2^{nd}$ jet 1 ms after el. $SOI_3$. In contrast to the initially small and then gradually growing HT ignition kernel development of OP1, a large HT reaction zone appears centered around that area. The HT reaction zone rapidly grows in size and intensity as it consumes the upstream mixtures and, unlike OP1, remains in a fixed location for several time steps. When transported downstream with the flow, the size of the reaction zone decreases slightly but with still growing intensity. Similar to OP1, the incomplete fuel consumption during HT combustion leads to post-combustion LT reactivity developing throughout the visible area.

In case of OP3, both LT and HT ignition deviate significantly from the previously described behavior of OPs 1 and 2. As a reminder: Due to the increased charge pressure, OP3 has the lowest global equivalence and compression ratio, the latter resulting in reduced compression temperatures per time step. The latter leads to a significant delay of LT reactivity of the $2^{nd}$ injection until 2 ms after el. $SOI_2$. Unlike OPs 1 and 2, where first LT oxidation reactions are located in the downstream half of the $2^{nd}$ jet, the initial LT reactivity of OP3 appears at the tip of the $2^{nd}$ spray at an axial distance between 50 and 60 mm from the injector. This is due to the fact that, as soon as ambient temperatures are sufficiently high to trigger LT reactivity, the upstream mixing is further advanced so that mixtures are too lean, while equivalence ratios at the spray head have reached a level suitable to promote LT oxidation reactions. In the following, LT reactivity spreads in the upstream half of the $2^{nd}$ spray, but without HT ignition to occur. In contrast to OPs 1 and 2, where the incoming 3rd injection immediately undergoes LT oxidation reactions, the absence of LT combustion products from the 2nd fuel spray fails to trigger LT reactivity inside the boundaries of the 3rd jet (small spots from 0.5 to 1 ms after el. SOI$_3$ are most likely background artifacts). The 3rd fuel spray eventually undergoes LT oxidation reactions 1.1 ms after el. SOI$_3$ and merges with the still persisting LT reaction zone of the 2nd jet. The onset of HT combustion then occurs 1.4 ms after el. SOI$_3$ at the head of the 3rd fuel spray and from there, the reaction zone quickly expands in the upstream direction. The resulting HT reaction is less intense compared to OPs 1 and 2 and its location is less persistent, so that it rapidly extinguishes when transported downstream with the flow, leaving a reduced LT reactivity in its wake. Both the decreased intensity and duration of the HT combustion event as well as the lowered pre- and post-combustion LT reactivity explain the very low MFB of OP3 with 39 % of the injected fuel mass only.

OP4 in comparison to OP1 faces the same amount of total fuel injected and is operated with the same charge. However, the timing of the 1st injection is delayed. This leads to a different mixing pattern. According Fig. 10 left, only 25 % of the light sheet plane are filled with fuel-air mixture prior to the 2nd injection. However, these 25% are more fuel rich because the 1st fuel spray has less time to mix (Fig. 10 right). This means that less fuel is entrained by the 2nd jet in the upstream regions of the spray, thereby leading to lower equivalence ratios in the upstream regions. The lower equivalence ratio in comparison to OP1 in this region is not expected to influence the local temperature due to the comparably large dwell between 1st and 2nd injection. With similar local temperature fields and richer conditions, the LT reactions of OP1 are occurring earlier and further upstream in comparison to OP4. Similar to OP3, first LT oxidation reactions are, therefore, shifted to the head of the 2nd spray, however, occur already 1.6 ms after el. SOI$_2$. Since the compression ratio of OP4 is considerably higher than that of OP3, the resulting increase in ambient temperature promotes LT reactivity at the comparably rich spray head region at a much earlier stage, while the upstream portions of the spray are still too lean to ignite. In the following, LT reactions are restricted to the head of the 2nd spray and do not spread further upstream, until auto-ignition occurs at a similar location 2 ms after el. SOI$_2$. The resulting reaction zone consumes the combustible mixture in the spray head region and extinguishes quickly 2.4 ms after el. SOI$_2$. Even though LT combustion products are absent in the near nozzle region, unlike OP3, high temperatures due to the elevated compression ratio and the already started HT combustion of the 2nd jet provoke LT reactivity within the limits of the 3rd jet immediately after it becomes visible. This is quickly followed by two auto-ignition kernels emerging between 30 and 40 mm downstream from the nozzle, which gradually grow into a single expanding reaction zone. The then progressing HT combustion event is very similar in intensity and duration to OP1 and eventually leaves a post-combustion formaldehyde signal in the wake of the downstream transported HT reaction zone.

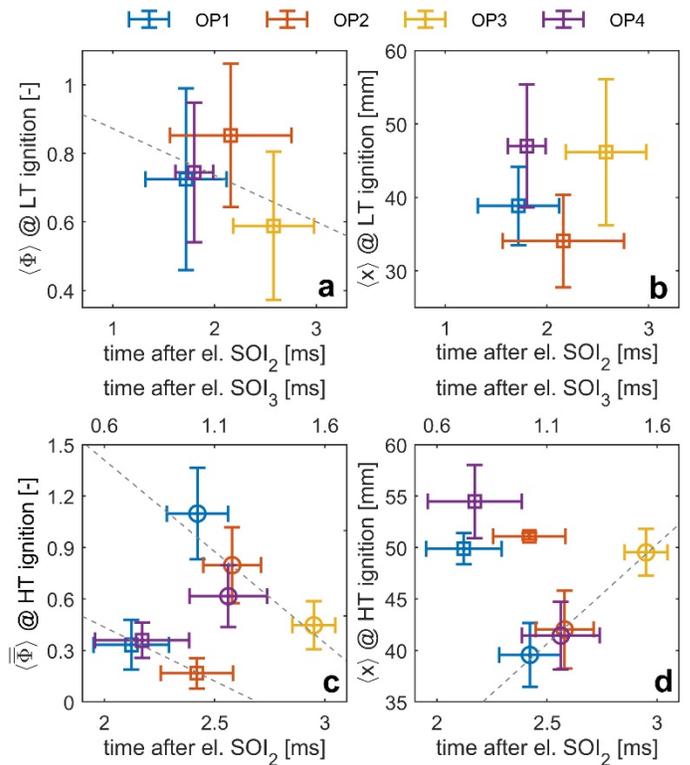

Fig. 14: Ignition statistics: Spatially averaged (a) equivalence ratio and (b) axial distance from the injector at LT ignition locations plotted against LT ignition delay. (c) Cross-sectional average equivalence ratio and (d) axial distance from the injector at HT ignition locations plotted against HT ignition delay related to SOI 2nd (bottom x-axis) and 3rd (top x-axis) injection. Symbols □ and ○ are used to discriminate between ignition events related to 2nd or 3rd injection, errorbars indicate the spread of the corresponding variable over the ten repetitions. Dashed grey lines show respective trends and are inserted to orient the reader.

To further substantiate above findings, LT and HT ignition timings as well as ignition locations are analyzed for all measurement runs and combined with respective results of the ensemble averaged mixing field. Although rising temperatures with respect to LT fuel oxidation will lead to local expansion of the flow field and thus change the local equivalence ratio distribution, it is argued that the following results nevertheless will display the qualitative trends in relation to the operating strategy. Fig. 14 (a) and (b) show equivalence ratios averaged over the initial LT ignition zones and their mean axial distance from the injector plotted against the LT ignition timing. Average equivalence ratios at LT ignition are all on similar levels and, with corresponding values for OPs 1 and 4 as well as slightly higher and lower magnitudes for OPs 2 and 3, follow the operating strategy. Likewise, the LT ignition timings are dominated by the varying compression ratio and the associated differences in ambient temperature, so that OPs 1 and 4 have a similar LT ignition delay, while LT oxidation reactions of OPs 2 and 3 are delayed. The average axial distances of the LT ignition event from the injector nozzle in Fig. 14 (b) confirm the trends found in the time-resolved analysis, whereby OPs 1 and 2 start LT oxidation reactions between 30 and 40 mm and OPs 3 and 4 further downstream around 50 mm. This distinct behavior is associated with the local structure of the mixing field: Even though differences in equivalence ratio at the location of LT ignition are small, the entrainment of base load fuel by the 2nd injection leads to richer fuel-air mixtures, so that in case of OPs 1 and 2 the ignition location moves further upstream where mixing conditions are favorable to support LT oxidation reactions. The 2nd fuel spray of OP3 entrains base load mixture as well, however, mixtures in the upstream part of the jet remain too lean so support LT reactivity, so that the LT ignition is shifted to the richer spray head region. Regarding OP4, the absence of base load fuel due to the late 1st injection results in leaner conditions within the boundaries of the 2nd spray. Therefore, the equivalence ratio in the upstream jet is too low to trigger LT oxidation reactions, so that the initial LT reactivity occurs at the richer spray head.

A similar analysis is performed for HT ignition in Fig. 14 (c) and (d). In order to link HT ignition timings to the local mixing state, planar equivalence ratio fields measured by tracer PLIF cannot be used right away, since the depth position of the in-plane fuel-air distribution does not necessarily correspond to that of an ignition kernel in the line-of-sight integrated $OH^*$ chemiluminescence signal. However, as the axial coordinates of $OH^*$ chemiluminescence and mixing field are the same, cross-sectional average equivalence ratios can be used for that purpose instead. As discussed above, the HT ignition of OPs 1, 2 and 4 can be separated in two distinct events related to the arrival of the 2nd and the 3rd fuel spray. This is indicated by the square (2nd) and circle (3rd) symbols and the upper and lower time axis in relation to $SOI_2$ and $SOI_3$, respectively. In case of OP3, HT ignition is characterized by a single ignition event only. When comparing ignition delay and local mixing state between the two HT events for OPs 1, 2 and 4, the first ignition occurs at very lean conditions with ignition delays twice as long, which is well in line with previous findings in the literature [21, 23]. It is visible, that the HT reactions in contrast to the LT reactions show a stronger dependency on the equivalence ratio, which is in agreement with previous works for mixtures of hydrocarbon fuels with air [24]. In particular, this is visible in the second HT ignition event, which is significantly influenced by flow and chemistry related mechanisms: OP1 has the shortest second HT ignition delay and ignites at the smallest axial distance. This behavior has its origin in an increased upstream LT reactivity within the 2nd fuel spray, which generates hot reaction products that are entrained by the expanding 3rd injection, combined with a relatively high compression ratio and, thus, ambient temperature. In terms of upstream LT reactivity, the 3rd injection of OP2 experiences similar conditions compared to OP1. However, as ambient temperatures are reduced due to the lower compression ratio, the second HT ignition event is slightly delayed and is located further downstream in a leaner mixing state. OP4 on the other hand has similar ambient conditions compared to OP1, but lacks the upstream LT reactivity of the 2nd fuel spray, leading to a similar second HT ignition delay, position and mixing state compared to OP2. Finally, due to the low compression ratio combined with the leanest fuel-air mixtures, HT ignition of OP4 is significantly delayed and shifted downstream. As indicated by the lowest MFB of 39 %, significant portions of the fuel-air mixture stay below the flammability limit.

# 4 Conclusion

In this study, mixing, ignition and combustion behavior in an optically accessible RCEM operated under PCCI relevant conditions were investigated by combined passive optical and laser-optical high-speed diagnostics. The Schlieren technique was applied to extract fuel spray contours and jet metrics. In a reactive campaign, this was combined with $CH_2O$ PLIF and $OH^*$ chemiluminescence imaging to detect LT and HT ignition delay and location and to further characterize combustion. Corresponding equivalence ratio fields were obtained from tracer PLIF measurements under inert conditions. In order to achieve the balance between fuel-air mixture stratification and premixing that is required for the PCCI combustion mode, a split injection schedule consisting of a first long base load injection early in the compression stroke and two short injections close to TDC was used. The applied operating strategy resulted in only minor changes in engine metrics such as global equivalence ratio or compression ratio. Nevertheless, distinct differences in local fuel-air mixing and ignition behavior as well as combustion efficiency were observed. In focusing on the short double injection, the analysis of the optical data revealed several key findings:

1. Previous experimental and numerical studies of closely coupled double injections under constant ambient conditions showed an increased penetration rate of the subsequent fuel spray caused by the persistent momentum of the preceding jet [20, 21, 47]. Under transient RCEM conditions, however, the aerodynamic gain from the preceding injection is counteracted by the density rise during the compression stroke, eventually leading to shorter penetrations of the consecutive jet.
2. In a double injection scheme, mixing of the subsequent fuel spray benefits from the turbulence induced by the preceding jet [13, 20, 21]. Although the shorter penetration of the subsequent jet during RCEM operation is confirmed by the equivalence distribution, an enhanced mixing rate of the second fuel spray is found for all investigated conditions. It is shown that, in contrast to the typical stagnation plane structure at the head of the preceding spray, the consecutive jet exhibits a significantly smoother gradient of the equivalence ratio at the spray tip, indicating faster mixing [20]. This is further quantified for all OPs by comparing the integrated cross-sectional average equivalence ratio at the spray head and tail of preceding and subsequent injection, revealing consistently faster mixing for the consecutive jet with an almost constant offset in the tail region, but substantial differences at the spray head.

3. The analysis of the mixing state within the laser light sheet after the first base load injection exhibits substantial differences resulting from the operating strategy. Although these marked differences in premixing are observed, all OPs approach a similar mixing state at the end of the following double injection event. Therefore, it is concluded that the in-plane mixing state is not determined by the operating strategy but dominated by the double injection event, the latter being identical for all OPs.

4. Even though variations in operating parameters are small, the thermodynamic analysis reveals a distinctly different combustion behavior between the OPs. All HRR curves display a gradual increase associated with the LT fuel oxidation, followed by a steep rise as a consequence of HT combustion. Since the mass fraction burned is significantly below 100 % in all OPs, three out of four HRR curves display post-combustion heat release by a persistent LT reactivity. From a further analysis of ignition timings it is found that more than 20 % of the overall heat release can be attributed to LT oxidation reactions before the actual onset of HT combustion.

5. The analysis of LT ignition characteristics reflects the complex interplay between mixing dynamics and reaction kinetics. The onset of LT reactivity after the $2^{nd}$ injection is found to be dominated by the operating strategy and the resulting variation in bulk temperature. Regarding the location of LT ignition, however, a distinct relationship to the local mixing state is revealed: Whereas the entrainment of fuel mass from the $1^{st}$ injection by the $2^{nd}$ jet leads to richer upstream mixtures and thus favors the onset of LT oxidation reactions, the absence of base load fuel in the near nozzle area results in fast leaning of the spray tail, so that LT ignition is shifted towards the richer spray head. An immediate onset of LT reactivity is observed when the $3^{rd}$ fuel spray expands into hot LT combustion products from the preceding $2^{nd}$ jet [20, 21]. In the absence of prevailing LT oxidation reactions, LT reactivity of the $3^{rd}$ jet is triggered by suitable ambient conditions. In addition, aerodynamic effects induced by the sustained momentum of the $2^{nd}$ jet promote the transition of the mixing field at the spray head of the $3^{rd}$ injection from a stagnation plane structure to a smoothly decaying equivalence ratio distribution and, thereby, further accelerate the onset of LT reactivity.

6. The HT ignition process is observed to be significantly influenced by the local mixing state in combination with the interaction between aerodynamics and combustion chemistry. Two distinct HT ignition events are identified for OPs 1, 2 and 4, which are associated with the arrival of the $2^{nd}$ and the $3^{rd}$ fuel spray, respectively. When comparing the two ignition events, the first occurs at much leaner conditions with ignition delays twice as long and at a notably larger axial distance from the injector nozzle. Regarding the second ignition event, it is shown that, even though similar ambient conditions prevail, the absence of LT combustion products in the upstream region leads to a prolonged HT ignition delay. In contrast, the entrainment of persistent upstream LT reactivity by the subsequent fuel spray results in shorter HT ignition delays and smaller axial distances of the initial reaction zone from the injector.

7. A sustained $CH_2O$ PLIF signal is observed in all OPs after HT combustion has finished, which is related to post-combustion LT oxidation reactions of the unburned fuel mass [21]. Compared to low signal levels prior to HT combustion, post-combustion fluorescence intensities are much higher, which, however, do not reflect the contribution of post-combustion LT reactivity to the overall heat release. It is argued, that even though the high temperature and pressure environment after HT combustion should promote quenching of formaldehyde fluorescence, this is overcompensated by the low oxygen content of the burnt gas. However, the quenching rate of formaldehyde fluorescence by oxygen under high temperature and pressure conditions has to be considered largely unknown, so that investigations in that direction would be highly desirable.

Above findings provide a conclusive analysis of mixing, ignition and combustion characteristics related to PCCI operation using a split injection schedule under transient engine conditions. In summary, the study provides clear evidence that even small changes in the operating strategy have significant influence on the local mixing state and reaction kinetics, which ultimately develop a huge impact on the overall engine efficiency. Future work will focus on CFD simulations of the experiments presented in this work, where current results provide an excellent validation database as well as a starting point for a further in-depth analysis of the complex coupling between global engine parameters and their impact on local mixture preparation, combustion properties and flow field.

# Acknowledgements

Funding from FVV (Forschungsvereinigung Verbrennungskraftmaschinen | Research Association for Combustion Engines, project no. 6013522) and from the Swiss Federal Office of Energy (grant no. Sl/501744-01) is gratefully acknowledged.

# Abbreviations

| | | | |
|---|---|---|---|
| BDC | Bottom Dead Center | OP | Operating Point |
| CFD | Computational Fluid Dynamics | PCCI | Premixed Charge Compression Ignition |
| CVC | Constant Volume Cell | PDF | Probability Density Function |
| EGR | Exhaust Gas Recirculation | PLIF | Planar Laser Induced Fluorescence |
| HCCI | Homogeneous Charge Compression Ignition | RCEM | Rapid Compression and Expansion Machine |
| HRR | Heat Release Rate | rms | root mean square |
| HT | High Temperature | SOI | Start Of Injection |
| LT | Low Temperature | TDC | Top Dead Center |
| LTC | Low-Temperature Combustion | UHC | Unburned Hydrocarbons |
| MFB | Mass of Fuel Burnt | | |

## List of Symbols

| | |
|---|---|
| $\Phi, \langle\Phi\rangle, \bar{\bar{\Phi}}$ | Equivalence ratio, ensemble averaged, cross-sectional average |
| $\rho_{air}$ | Air density [kg/m$^3$] |
| $\rho_a$ | Charge density [kg/m$^3$] |
| $\phi$ | TMPD fluorescence quantum yield |
| $AFR_{stoich}$ | Stoichiometric air-fuel-ratio for n-heptane-air mixtures |
| $I$ | Background corrected CH$_2$O-PLIF intensity |
| $I_0$ | Incident laser energy |
| $I_{ff}$ | Flat-field image |
| $K$ | PLIF-intensity to fuel density conversion factor |
| $P$ | Probability |
| $P_a$ | Charge pressure [Pa] |
| $P_f$ | Fuel pressure [Pa] |
| $S$ | Spray penetration length [mm] |
| $T$ | Temperature [K] |
| $x, r$ | Axial, radial coordinate [mm] |
| $t$ | Time after start trigger [ms] |
| $a_0$ | Spray model calibration parameter |
| $d_0$ | Nozzle reference diameter [mm] |
| $p$ | Pressure [Pa] |